\documentclass{iaus}
\usepackage{graphicx}

\title[PopIII Core Collapse Supernovae] 
{Nucleosynthesis of PopIII Core Collapse Supernovae and the abundances in extremely metal poor stars}

\author[Limongi \& Chieffi]   
{M. Limongi$^1$ \and A. Chieffi$^2$}

\affiliation{$^1$INAF - Osservatorio Astronomico di Roma, Via Frascati 33, I-00040 Monteporzio Catone (Roma), ITALY \break email: marco@mporzio.astro.it\\[\affilskip]
$^2$INAF - Istituto di Astrofisica Spaziale e Fisica Cosmica, Via Fosso del Cavaliere, I-00133, Roma, ITALY \break email: achieffi@rm.iasf.cnr.it}

\pubyear{2005}
\volume{228}  
\pagerange{1--7}
\date{?? and in revised form ??}
\jname{From Lithium to Uranium: Elemental Tracers of Early Cosmic Evolution}
\editors{V. Hill, P. Fran\c{c}ois \& F. Primas, eds.}
\begin{document}

\maketitle

\begin{abstract}
We present a new analysis of the abundances observed in extremely metal
poor stars based on both a new generation of theoretical presupernova models and explosions of zero
metallicity massive stars and a new abundance analysis of an homogeneous sample of stars having
$\rm [Fe/H]\leq -2.5$ (\cite[Cayrel et al. 2004]{cayrel04}).
\keywords{nucleosynthesis, stars:abundances, supernovae:general}
\end{abstract}

\firstsection 
\section{Introduction}
Extremely metal poor stars, i.e., those stars having $\rm [Fe/H]\leq -3.0$, formed
in the very early epochs of Galaxy formation hence they reflect the ejecta
of the first core collapse supernovae. Weather or not they are associated to single
supernovae (\cite[Ryan, Norris \& Beers 1996]{rnb96}, \cite[Audouze \& Silk 1995]{as95}) or single
burst events (\cite[Cayrel et al. 2004]{cayrel04}) they provide very useful constraints to test presupernova
models, supernova explosions and nucleosynthesis theories. Moreover the element abundance
patterns observed in these stars can be used to infer the nature of the first generations of
stars and supernovae.

In a recent paper (\cite[Chieffi \& Limongi 2002]{cl02}) we presented a detailed comparison 
of an extended set of elemental abundances observed in some of the most metal poor stars known
at that time and the ejecta produced by a generation of primordial core collapse supernovae.
In particular we defined a "template" ultra-metal poor star that represented all the stars in the
database and that was compared to the theoretical predictions. Our main findings on that paper were
the following: (1) the fit to the "template" star, in particular the [Si/Mg] and [Ca/Mg] ratios,
favored a large carbon abundance at the end of central
He burning that, in the framework of classical convection theory, implied a low $\rm ^{12}C(\alpha,\gamma)^{16}O$
cross section - in this case we obtained a good fit to the 8 of the 11 observed element abundance ratios; 
(2) the fit to [Sc/Fe] and [Co/Fe] drastically depended on the central C abundance left by core He burning - 
at variance with the current beliefs that it is difficult to interpret the observed large overabundance of [Co/Fe],
a mildly large carbon abundance in the He exhausted core allowed a very good fit to [Co/Fe]; (3)
within the grid of models computed at that time it was not possible to find a simultaneous good fit
to the remaining three elements, Ti, Cr and Ni, even for an arbitrary choice of the mass cut.

The aim of this paper is to present a new analysis of the abundances observed in extremely metal
poor stars based on both a new generation of theoretical presupernova models and explosions of zero
metallicity massive stars and a new abundance analysis of an homogeneous sample of stars having
$\rm [Fe/H]\leq -2.5$ (\cite[Cayrel et al. 2004]{cayrel04}).

\section{Theoretical models}
The analysis presented in this paper is based on a new generation of presupernova models and explosions
of zero metallicity massive stars. These models, that will be discussed in more detail in a forthcoming
paper (\cite[Chieffi \& Limongi 2005]{cl05}), have been computed by the latest release (5.050218) of the FRANEC (Frascati 
RAphson Newton Evolutionary Code) - the previous version has been presented by \cite{lc03}, and references therein. 

The main differences with respect to our previous set of models (\cite[Chieffi \& Limongi 2002]{cl02}) 
are the following: the nuclear network is much more extended w.r.t. the previous one and includes now 
282 nuclear species, from H to $\rm ^{98}Mo$, linked by about 3000 reactions; the convective 
mixing is performed by solving a diffusion equation rather than adopting the \cite{se80} algorithm 
(see \cite[Chieffi, Limongi \& Straniero 1998]{cls98}); the mixing and nuclear burning 
are fully coupled together and solved simultaneously; updated cross sections have been used whenever possible 
(see \cite[Chieffi \& Limongi 2005]{cl05} for a full list of references), here we just mention that the
cross section for the $\rm ^{12}C(\alpha,\gamma)^{16}O$ is taken from the latest analysis performed by
\cite{kunz02}.

Also the computation of the explosive nucleosynthesis has been greatly improved w.r.t. the previous technique (based
on the radiation dominated shock approximation). Indeed, in the present models the explosion has been induced
by means of a piston located within the iron core (typically ad $\rm \sim 1 M_\odot$) with a given initial velocity and then
the passage of the consequent blast wave through the mantle of the star has been followed by means of a 
PPM hydro code (also this code is presented in detail in \cite[Chieffi \& Limongi 2005]{cl05}).

\begin{figure}
\centerline{
\scalebox{0.4}{
\includegraphics{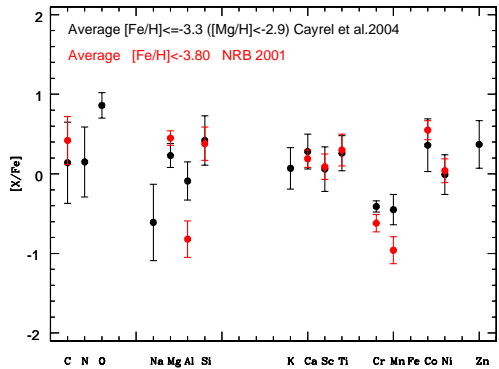}
}
}
\caption{Comparison between average star computed from the \cite{nrb01} database ({\em red dots})
and that obtained from the \cite{cayrel04} database ({\em black dots}).}
\end{figure}

\section{The observable}
The analysis presented in \cite{cl02} was based on the observations of five stars of metallicity lower than
[Fe/H]=-3.3 (\cite[Norris, Ryan \& Beers 2001]{nrb01}). Since three of the five stars, i.e., $\rm CD~38^{o}245$,
CD 22172-002 and CS 22885-096, showed a remarkably similar pattern we could define an "average" (AVG01) star that represented all three of them.
Today, a new abundance analysis of a homogeneous sample of 35 stars is available (\cite[Cayrel et al. 2004]{cayrel04}).
Within this new database, 22 of the 35 stars have $\rm [Fe/H]<-3.0$ and below [Mg/H]=-3.0 (which correspond
to [Fe/H]=-3.3) the observed element abundance ratios are remarkably uniform, suggesting that the level
of "primordial yields" may have been reached. As a consequence, as in \cite{cl02}, we can define an average
star (AVG04) that should represent all the stars having [Fe/H] lower than -3.3. A comparison between AVG01 and 
AVG04 (Fig.1) shows that: (1) more elements are observed for the AVG04 compared to AVG01; (2)  
AVG04 shows lower values of [C/Fe], [Mg/Fe] and [Co/Fe] and higher values of [Al/Fe], [Cr/Fe] and [Mn/Fe] compared to AVG01 -
also the relative scaling of these elements are obviously different between the two stars; (3) for each element 
AVG04 shows larger dispersions compared to AVG01 - this is due to the fact that the average abundances
for AVG04 have been obtained with a larger number of stars compared to AVG01.

\section{Fit to light elements (C-Ca)}
Since the elements from C to Ca are not affected by the exact location of the mass cut
(see \cite[Chieffi \& Limongi (2002)]{cl02} for a more details) and since we are
interested here in comparing their relative scaling with the observations, it is useful to
normalize all their abundances to one of them. We choose to normalize all the abundances
of all these elements to Mg. Figure 2 shows the comparison between the AVG04 star and
the ejecta of all the models in our grid under the hypothesis
that all the models eject some amount of Fe.

\begin{figure}
\centerline{
\scalebox{0.45}{
\includegraphics{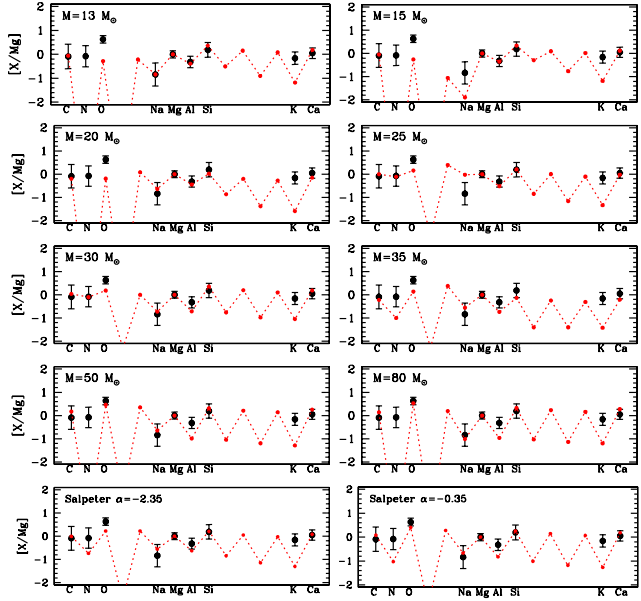}
}
}
\caption{Upper eight panels: comparison between the AVG04 star (black filled dots) and the ejecta of all the models in our grid
(red filled dots connected by a dotted line). Lower two panels: comparison between the AVG04 star and the ejecta 
of all the models in our grid averaged over a Salpeter IMF (red filled dots connected by a dotted line) 
with two slopes, i.e., $\alpha=-2.35$ and $\alpha=-0.35$.}
\end{figure}

The first upper eight panels of Figure 2 clearly shows that the fit to [C/Mg], [Na/Mg], [Si/Mg] and [Ca/Mg] is remarkably good for
almost all the models without any sizeable dependence on the initial mass. On the contrary the [O/Mg]
ratio would favor high progenitor masses while [Al/Mg] would be better reproduced by low progenitor masses.
The [N/Mg] log ratio is very well fitted only by stars in the mass range between 25 and $\rm 30~M_\odot$.
Such a strong primary N production is connected, in these stars, with the ingestion of protons by the He convective
shell the penetrates into the H rich layer. Finally, [K/Mg] is always significantly underestimated by all the models.
The two lower panels of Figure 2 shows the comparison between the AGV04 star and the ejecta provided by a generation of massive
stars (in the mass range 13-$80~M_\odot$) averaged over a Salpeter IMF for
two choices of the slope $\alpha$. In particular, in the standard case ($\alpha$=-2.35), where the low mass
massive stars dominate, N and O are both underproduced while Al is quite well fitted. 
An opposite behavior is found when the slope of the IMF is lower
($\alpha$=-0.35) and the relative contribution of the more massive stars is higher.
In any case N is significantly underestimated because the high primary N production occurs
in a very narrow mass interval around the $\rm 25~M_\odot$.

\begin{figure}
\centerline{
\scalebox{0.45}{
\includegraphics{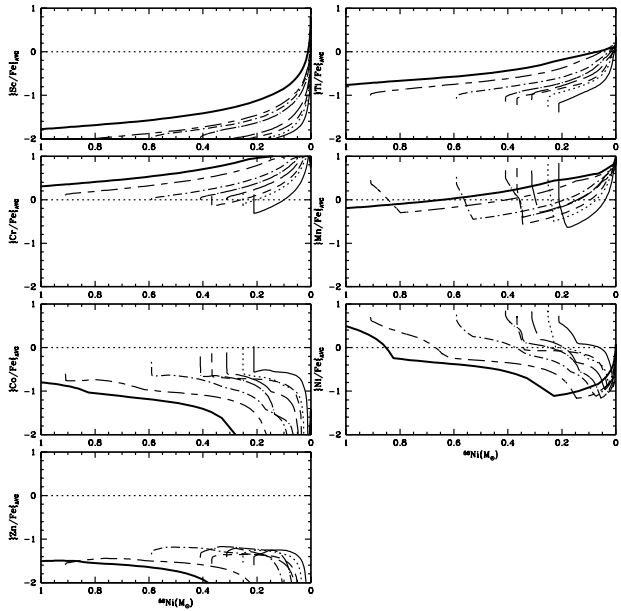}
}
}
\caption{Trends of all the elements produced by the explosive complete and/or incomplete Si burning
with the amount of Fe ejected. The eight lines in each panel refer to  13 ({\rm solid}), 
15 ({\rm dot}), 20 ({\rm short dash}), 25 ({\rm long dash}), 30 ({\rm dot-short dash}), 
35 ({\rm dot-long dash}), 50 ({\rm short dash-long dash}) and $\rm 80~M_\odot$ ({\rm tick solid}).} 
\end{figure}

\section{Fit to the heavy elements (Sc-Zn)}
All the elements between Sc and Zn are produced by either the complete and/or the incomplete
explosive Si burning in the deepest layers of the exploding envelope, hence they are those
mostly affected by the location of the mass cut.
In order to compare the theoretical yields of these elements with their corresponding observations
it is useful to introduce the following abundance ratio:
$\rm \{X/Fe\}_{star}=log(X/Fe)_{model}-log(X/Fe)_{star}$ (see \cite[Chieffi \& Limongi 2002]{cl02}).
This ratio is similar to the standard [X/Fe] but in this case the reference star
is not the Sun anymore but the star to be fitted. Obviously a perfect fit to the observations
is obtained when $\rm \{X/Fe\}_{star}=0$. Figure 3 shows how $\rm \{Sc,Ti,Cr,Mn,Co,Ni,Zn/Fe\}$ vary with the
amount of $\rm ^{56}Ni$ ejected (i.e., their dependence on the mass cut) for each stellar model.

From Figure 3 it is clear that there is no value of the mass cut that lead to a simultaneous
fit of all the heavy elements from Sc to Zn. In particular, 
there is no hope of obtaining a good fit to Sc, Ti, Co and Zn for any choice of the mass cut while,
on the contrary, there exists a value of $\rm ^{56}Ni$ that would allow a fit to Cr, Mn and Ni.
In this last case, however, only Mn and Ni can be simultaneously fitted by choosing the same
value of the mass cut.

\begin{figure}
\centerline{
\scalebox{0.45}{
\includegraphics{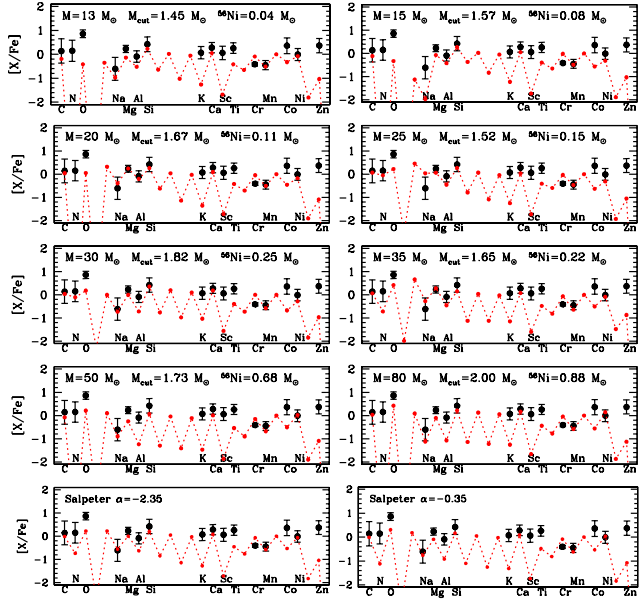}
}
}
\caption{Same as Fig.2 but with a choice of the mass cut (or equivalently $\rm ^{56}Ni$), shown in each
panel, that provides the best fit to the heavy elements.} 
\end{figure}

Figure 4 shows the best overall fit of all the elements for each stellar model (first upper eight panels)
and the best fit obtained by integrating the yields of each model over a Salpeter IMF (two lower panels).
Since the amount of Fe required to obtain the best fit to the heavy elements is slightly larger
(i.e., a deeper mass cut) than the one needed to fit the light elements, the block of the light
elements lowers somewhat without, however, changing their relative scaling (see above).
We have shown the best overall fit of all the elements just for sake of completeness, but
it is clear that the main problem in the comparison between theory and observations remain the
heavy elements Ti, Sc, Cr, Co and Zn. 

While we have no idea about the discrepancy in [Ti/Fe] and [Cr/Fe], we guess that 
the fit to Sc, Co and Zn could be significantly improved by adopting a slightly lower
$\rm ^{12}C(\alpha,\gamma)^{16}O$ cross section (still in the range of the experimental uncertainties),
as we already have shown in \cite{cl02}. 
We plan to investigate such a possibility in a forthcoming paper.

An alternative solution that would improve the fit to the heavy elements
has been recently proposed by \cite{UN05}. In particular \cite{UN05} claimed that the overall fit to the heavy elements
is greatly improved by means of a more energetic explosions, by assuming a mixing-fall back mechanism,
by arbitrarily imposing $Y_{e}=0.5001$ in the explosive complete Si burning region and finally 
by reducing the density of the presupernova model in the innermost zones by a factor of $\sim 3$ w.r.t.
the original one. Although a quite good fit can be obtained by arbitrarily changing the properties of
the presupernova model and by assuming a more energetic explosion, we like to think that
few crucial quantities (like, e.g., the $\rm ^{12}C(\alpha,\gamma)^{16}O$ cross section) can
drive the theory towards observations.

\section{Conclusions}
We presented a new comparison between the element abundance ratios observed in extremely metal
poor stars ($\rm [Fe/H]<-3.3$, \cite[Cayrel et al. 2004]{cayrel04}) and our
latest yields of zero metallicity core collapse supernovae.

Our main findings are the following: (1) The elements that do not depend on
the location of the mass cut (from C to Ca) are quite well reproduced by the models, in particular:
[C,Na,Si,Ca/Mg] fit pretty well the observations without any significant dependence
on the progenitor mass; [O/Mg] and [Al/Mg] show an anti correlated trend, i.e., 
[O/Fe] would favor high mass progenitors while [Al/Mg] is better reproduced by
lower mass progenitors; [N/Mg] is very well fitted only by stars in the mass range
$\rm 25-30~M_\odot$, the high primary N production in these stars being the result of
the He convective shell penetration into the H rich layers; [K/Mg] is always significantly
underproduced by all the models in the grid. (2) The elements that depend on the location
of the mass cut (from Sc to Zn) are never simultaneously reproduced by the models for
any (unique) choice of the mass cut, [Mn/Fe] and [Ni/Fe] are the only ones that could
be simultaneously fitted for the same value of the mass cut.
While we do not have any idea about the discrepancy in [Ti/Fe] and [Cr/Fe], we suggest that the
adoption of a slightly lower $\rm ^{12}C(\alpha,\gamma)^{16}O$ cross section 
(still in the range of the experimental uncertainties) could significantly improve
the fit to the observations of Sc, Co and Zn. We will investigate such a possibility
in a forthcoming paper.

\begin{acknowledgments}
I (ML) would like to thank Tatiana Grilli for her continuous encouragement and support
during the very hard last year of my life.
\end{acknowledgments}


\begin{thebibliography}{}

\bibitem[Audouze \& Silk (1995)]{as95}
     {Audouze, J., Silk, J.} 1995,  
     \textit{ApJ} (Letters) {451}, L49

\bibitem[Cayrel \etal\ (2004)]{cayrel04}
     {Cayrel, R., Depagne, E., Spite, M., Hill, V., Spite, F., Fancois, P., Plez, B.,
      Beers, T., Primas, F., Andersen, J., Barbuy, B., Bonifacio, P., Molaro, P., Nordstr\"om, B.} 2004,  
     \textit{A\&A} {416}, 1117

\bibitem[Chieffi, Limongi \& Straniero (1998)]{cls98}
     {Chieffi, A., Limongi, M., Straniero, O.} 1998,  
     \textit{ApJ} {502}, 737

\bibitem[Chieffi \& Limongi (2002)]{cl02}
     {Chieffi, A., Limongi, M.} 2002,  
     \textit{ApJ}, {577}, 281 

\bibitem[Chieffi \& Limongi (2005)]{cl05}
     {Chieffi, A., Limongi, M.} 2005,  
     \textit{in preparation}

\bibitem[Kunz \etal\ (2002)]{kunz02}
     {Kunz, R., Fey, M., Jaeger, M., Mayer, A., Hammer, J.W., Staudt, G., Harissopulos, S., Paradellis, T.} 2002,  
     \textit{ApJ} {567}, 643

\bibitem[Limongi \& Chieffi (2002)]{lc02}
     {Limongi, M., Chieffi, A.} 2002,  
     \textit{PASA} {19}, 1

\bibitem[Limongi \& Chieffi (2003)]{lc03}
     {Limongi, M., Chieffi, A.} 2003,  
     \textit{ApJ} {592}, 404

\bibitem[Norris, Ryan \& Beers (2001)]{nrb01}
     {Norris, J.E., Ryan, S.G., Beers, T.C.} 2001,
     \textit{ApJ} {561}, 1034

\bibitem[Ryan, Norris \& Beers (1996)]{rnb96}
     {Ryan, S.G., Norris, J.E., Beers, T.C.} 1996,
     \textit{ApJ} {471}, 254

\bibitem[Sparks \& Endal (1980)]{se80}
     {Sparks, W.M., Endal, A.S.} 1980,  
     \textit{ApJ} {237}, 130

\bibitem[Umeda \& Nomoto (2005)]{UN05}
     {Umeda, H., Nomoto, K.} 2005,  
     \textit{ApJ}, {619}, 427 

\end{thebibliography}
\end{document}